\documentclass[aps,twocolumn,showpacs,superscriptaddress,10pt,reprint]{revtex4-2}

\usepackage{bm}
\usepackage{bbold}
\usepackage{graphicx}
\usepackage{svg}
\usepackage{color}
\usepackage[binary,amssymb]{SIunits}
\usepackage{comment}
\usepackage{hyperref}

\usepackage{amsmath,amssymb}
\usepackage{color}
\definecolor{Blue}{rgb}{0.3,0.3,0.9}
\definecolor{Red}{rgb}{0.9,0.3,0.3}
\definecolor{Green}{rgb}{0.3,0.6,0.3}

\begin{document}

\title[Bound states in the continuum in InSb nanowire networks]
{Uncovering bound states in the continuum in InSb nanowire networks}

\author{D. Mart\'{\i}nez}

\affiliation{GISC, Departamento de F\'{\i}sica de Materiales, Universidad 
Complutense, E-28040 Madrid, Spain}

\author{P. A. Orellana}

\affiliation{Departamento de F\'isica, Universidad T\'ecnica Federico Santa Mar\'{\i}a, Casilla 110 V, Valpara\'{\i}so, Chile}

\author{L. Rosales}

\affiliation{Departamento de F\'isica, Universidad T\'ecnica Federico Santa Mar\'{\i}a, Casilla 110 V, Valpara\'{\i}so, Chile}

\author{J. Dolado}

\affiliation{ESRF, The European Synchrotron, 71 Avenue des Martyrs, Grenoble 38043, France}

\author{M. Amado}

\affiliation{Nanotechnology Group, USAL-Nanolab, Universidad de Salamanca, E-37008 Salamanca, Spain}

\author{E. Diez}

\affiliation{Nanotechnology Group, USAL-Nanolab, Universidad de Salamanca, E-37008 Salamanca, Spain}

\author{F. Dom\'{i}nguez-Adame}

\affiliation{GISC, Departamento de F\'{\i}sica de Materiales, Universidad Complutense, E-28040 Madrid, Spain}

\author{R. P. A. Lima}

\affiliation{GISC, Departamento de F\'{\i}sica de Materiales, Universidad Complutense, E-28040 Madrid, Spain}
\affiliation{GFTC, Instituto de F\'{\i}sica, Universidade Federal de Alagoas, Macei\'{o} AL 57072-970, Brazil}
\affiliation{GISC, Departamento de Electrónica F\'{\i}sica, Ingenier\'{\i}a El\'ectrica y F\'{\i}sica Aplicada, ETSIT, Universidad Polit\'ecnica de Madrid, Avenida Complutense 30, E-28040 Madrid, Spain}

\begin{abstract}

Bound states in the continuum (BICs) are exotic, localized states even though their energy lies in the continuum spectra. Since its discovery in 1929, the quest to unveil these exotic states in charge transport experiments remains an active pursuit in condensed matter physics. Here, we study charge transport in InSb nanowire networks in the ballistic regime and subject to a perpendicular magnetic field as ideal candidates to observe and control the appearance of BICs. We find that BICs reveal themselves as distinctive resonances or antiresonances in the conductance by varying the applied magnetic field and the Fermi energy. We systematically consider different lead connections in hashtag-like nanowire networks, finding the optimal configuration that enhances the features associated with the emergence of BICs. Finally, the investigation focuses on the effect of the Rashba spin-orbit interaction of InSb on the occurrence of BICs in nanowire networks. While the interaction generally plays a detrimental role in the signatures of the BICs in the conductance of the nanowire networks, it opens the possibility to operate these nanostructures as spin filters for spintronics. We believe that this work could pave the way for the unambiguous observation of BICs in charge transport experiments and for the development of advanced spintronic devices.

\end{abstract}

\date{\today}

\maketitle

\section{Introduction}

Bound states in the continuum~(BICs) are spatially localized states or waves whose energy or frequency lies within a continuum spectrum of propagating modes. Von Neumann and Wigner proposed the existence of these exotic and counterintuitive states in the earlier days of quantum mechanics~\cite{Neumann1929}. They constructed a spatially oscillating attractive potential and solved the corresponding Schr\"{o}dinger equation, finding a truly localized state above the potential barrier as a result of destructive interference.  Much later, Stillinger and Herrick reexamined and extended these ideas in the context of atoms and molecules~\cite{Stillinger75}. 

The occurrence of BICs is related to the dynamics of coherent waves, and therefore, they have been thoroughly studied not only in atoms and molecules~\cite{Stillinger75,Friedrich85,Zhang12} but also in areas of optics and  photonics~\cite{Plotnik2011,Cerjan2021,Wang2021,Joseph2021,Shi2022,Liu2023}, plasmonics~\cite{Azzam2018,Liang2020,Sun2021}, acoustics~\cite{Deriy2021,Pu2023,Pan2023,Lee2023} and nanoelectronics~\cite{Ladron2006, Bulgakov2006, Gonzalez-Santander2013,Prodanovic2013,Mur-Petit2014,Grez2022}, to name a few (see Refs.~\cite{Hsu2016,Koshelev2023,Kang2023,Xu2023} for comprehensive reviews on BICs). Since the pioneering work by Plotnik \emph{et al.} on the symmetry-protected BICs in an array of parallel dielectric single-mode waveguides fabricated of fused silica~\cite{Plotnik2011}, much progress have been achieved in the observation and characterization of BICs in photonic structures~\cite{Marinica2008,Bulgakov2008,Bulgakov2010}. 

However, signatures of BICs in charge transport experiments remain elusive and less explored in the literature. Recent advances in nanotechnology have made it possible to conceive and fabricate quantum devices that support BICs. In this context, Albo~\emph{et al.} made use of intersubband photocurrent spectroscopy to demonstrate that a BIC exists above (Ga,In)(As,N) / (Al,Ga) as quantum wells that arises from the hybridization of nitrogen-related defect states and the extended states of the conduction band~\cite{Albo2012}. N\"{o}ckel investigated theoretically the ballistic electron transport across a quantum dot in a weak magnetic field~\cite{Nockel1992}. Resonances in the transmission were found to get narrower upon decreasing the magnetic field, signaling the occurrence of BICs as the magnetic field vanishes. Therefore, the external magnetic field can control the coupling of the spatially localized state with extended states of the continuum energy spectrum in nanostructures. In double or triple quantum dots, the coupling between the BICs and the continuum energy states can be controlled by detuning the energy levels of each quantum dot using gate voltages~\cite{Ladron2006,Gonzalez-Santander2013}, without the need of external magnetic fields. In this case, the appearance of BICs is identified by the occurrence of Fano resonances~\cite{Fano1961} in the transmission. 

In this work, we investigate the impact of BICs on the conductance of InSb nanowire networks in the ballistic regime, aiming to provide new routes for their observation in charge transport experiments. These nanostructures present phase-coherence lengths exceeding several micrometers with Aharonov–Bohm oscillations up to five harmonics due to high crystalline quality~\cite{OphetVeld2020}. In addition, InSb nanowires have been theorized to host Majorana zero modes~\cite{Oreg2010} and are regarded as suitable candidates for topological quantum computing~\cite{Stern2013}. Inspired by a study in which two interacting quantum dots exhibited BICs~\cite{Guevara2003}, our investigation focuses on nanowire networks forming a hashtag pattern. In this configuration, the corners emulate the quantum dots, while the branches correspond to tunneling processes between them. Here, the dead-end chains connected to the corners allow us to control the interference among various electron paths that help us to set the energy of the BICs. The main goal of this research is to determine the optimal configuration, including the placement of the leads and whether and where to place the dead-end chains, to enable the unambiguous observation of the BICs. Finally, the effect of the strong Rashba spin-orbit interaction (SOI) of InSb nanowires on the BICs and the associated spin currents is studied.  


\section{Model Hamiltonian and method}

\begin{figure*}
    \centering
    \includegraphics[width = 0.9\linewidth]{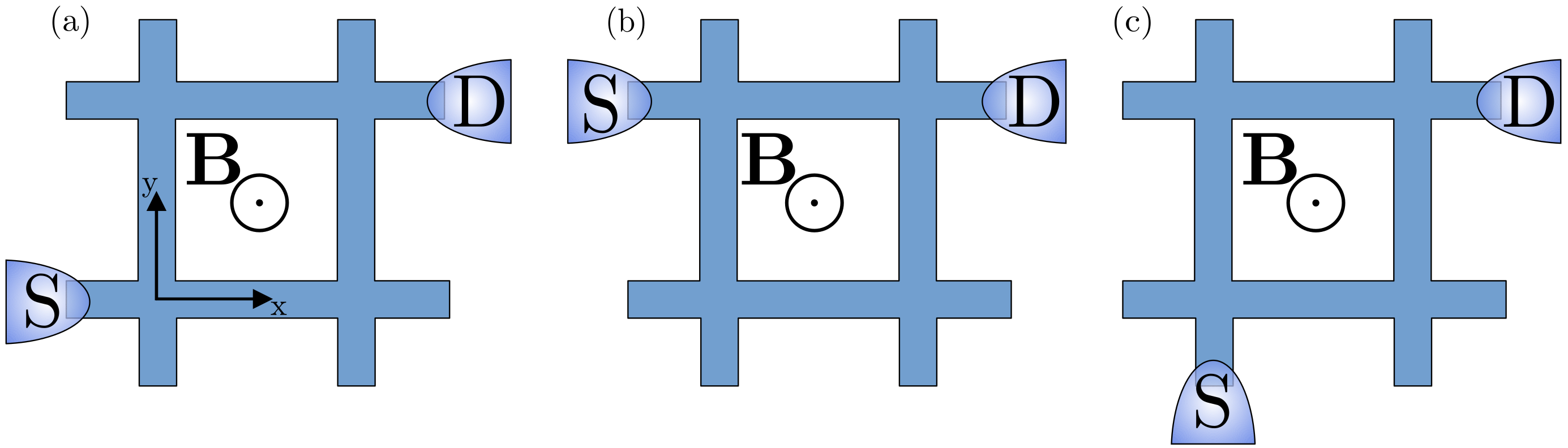}
    \caption{Sketch of the configurationally different ways to attach the leads onto the nanowire network. S and D represents the source and drain leads. All the structures are threaded by a perpendicular magnetic field.}
    \label{fig: 1}
\end{figure*}

We consider six different arrangements of leads and dead-end branches of the InSb nanowire network. All of them will have a hashtag structure, but different places where the leads are attached or the number of dead-end chains and where they are placed (see figures~\ref{fig: 1} and \ref{fig: 4}). Since the cross-section of the nanowires is small, the subbands are well separated in energy, and hence, the nanowires can be regarded as one-dimensional. By discretizing the Ben Daniel-Duke Hamiltonian, the resulting equation of motion for the envelope function becomes equivalent to a tight-binding model with nearest-neighbor coupling. The hopping energy is $t = \hbar^2/(2 m^* a^2)$, where  $m^*$ is the electron effective mass and $a$ is the grid spacing. In this framework, the single-electron Hamiltonian can be cast in the form $\mathcal{H}=\mathcal{H}_{\text{leads}} + \mathcal{H}_{\text{sys}} + \mathcal{H}_{\text{c}}$ where
\begin{subequations}
\begin{align}
\mathcal{H}_{\text{leads}}  = & t_l  \sigma_0 \sum_{l,j}  \Big( d_{l+1, j}^{\dagger} d_{l, j}^{} 
+ d_{l, j+1}^{\dagger} d_{l, j}^{}\Big) +\text{h.c.} \ , \\
\mathcal{H}_{\text{sys}}=&\sum_{n,m}\Big(\epsilon_{nm} \sigma_0 +g\mu_{\text{B}} B\sigma_z\Big) c_{n,m}^{\dagger}c_{n,m}^{}  \nonumber\\
-&\sum_{n,m}\left(t e^{-\mathrm{i} m a B} \sigma_0 -\frac{\mathrm{i}\alpha}{2} \sigma_y\right)c_{n+1,m}^{\dagger} c_{n,m}^{}\nonumber\\
-&\sum_{n,m}\left(t \sigma_0+\frac{\mathrm{i}\alpha}{2} \sigma_x\right)c_{n,m+1}^{\dagger} c_{n,m}^{}+\text{h.c.} 
\end{align}
\end{subequations}
Here, $d_{l,j} = (d_{l,j,\uparrow}, d_{l,j,\downarrow})^T$ is the annihilation operator of the leads at site $\{l, j\}$ of the lead while $c_{n,m} = (c_{n,m,\uparrow}, c_{n,m,\downarrow})^T$ is the annihilation operator of the nanowire networks at site $\{n, m\}$ of the hashtag pattern, $t$ and $t_l$ are the hopping energy in the network and the leads respectively, $\epsilon_{ij}$ is the site energy of the network taking the site energy of the leads as zero, $\alpha$ is the Rashba spin-orbit coupling, $\sigma_0$ is the $2\times2$ identity matrix and $\sigma_{x,y,z}$ are the Pauli matrices. The network is threatened by a magnetic field tuned to add a complex phase in hopping parameter along the $x$ direction. Finally, $\mathcal{H}_c$ describes the tunnel coupling between neighbor sites of the lead and the hashtag network, with hopping energy $t_l$ and a complex phase when the sites are aligned along the $x$ direction.

Since we are assuming that the phase coherence length is larger than the system size, the conductance of the device is calculated within the Landauer-B\"{u}ttiker formalism in the ballistic regime as follows
\begin{equation}
    G(E) = \frac{e^2}{h}\,\tau(E)\ ,
\end{equation}
where $\tau(E)$ is the transmission coefficient at the Fermi energy $E$ and $-e$ is the electron charge. Notice that we are neglecting the broadening of the Fermi-Dirac distribution function by taking $T=0\,$K. The transmission coefficient can be calculated straightforwardly with the aid of the transfer-matrix approach or the Green's functions method (see Supplementary Material). Additionally, when there is a spin-flip mechanism as Rashba SOI, the conductance can be obtained as
\begin{equation}
    G_{s,s'}(E) = \frac{e^2}{h}\, \tau_{s,s'}(E)\ ,
\end{equation}
where $s,s^{\prime}={\uparrow,\downarrow}$ stand for the spin projections of the incoming and outgoing electron states, respectively. Therefore, we can define the spin polarization of the conductances as follows
\begin{equation}
    P_S(E) = G_{\uparrow, \uparrow} + G_{\downarrow, \uparrow} - G_{\uparrow, \downarrow} - G_{\downarrow, \downarrow}\ .
\end{equation}

The BICs are difficult to find because they are not coupled to the continuum energy spectra until the symmetry that allows them to exist is broken. For this reason, we calculate the conductance by varying the magnetic field and the Fermi energy. When the magnetic field is switched on, the BICs can be uncovered as an emerging resonance or antiresonance in the conductance. Once the presumed BICs are found, their spatially localized nature is assessed through the participation ratio~(PR)
\begin{equation}
    \mathrm{PR} = \left( \sum_{n,m} \left| F_{nm}(E_\mathrm{BIC}) \right|^4 \right)^{-1}\ ,
\end{equation}
where $F_{nm}$ is the amplitude of the normalized envelope function at site $\{n,m\}$. Notice that $\mathrm{PR} = 1$ when the state is fully localized at a single site whereas $\mathrm{PR} \simeq N$ for an extended state.

\section{Results}

We calculate the conductance numerically with the Kwant toolkit~\cite{Groth2014}. The parameters considered for the system and leads are the same as in the previous section. Additionally, the length for the horizontal and vertical dead-ends are $l_x = 8\,$nm and $l_y = 10\,$nm, respectively. The InSb nanowires chosen for this study have an electron effective mass $m^*_\mathrm{InSb} = 0.014\,m_e$~\cite{Zawadzki1974} and gyromagnetic factor $g_\mathrm{InSb} \approx 50$~\cite{Nilsson2009}, where $m_e$ is the free electron mass. The leads are also assumed to be InSb. For the simulations, the dimensions of the hashtag patterns was $L = 50\,$nm for the length of the edges of the square. Finally, the grid spacing has been set to $a = 2\,$nm.

\subsection{Impact of the lead geometry}

We will consider three configurationally different ways to attach the leads. The first one connects the leads into two opposite corners along the $x$ direction. In the second one, the leads are connected one in front of the other along the $x$ direction. In the third one, the leads are connected at two opposite corners but in perpendicular directions (see figure~\ref{fig: 1}). The last geometry differs from the first one due to the effect of the magnetic field, which adds a phase difference between the two electron paths. 

\begin{figure*}
    \centering
    \includegraphics[width = 0.9\linewidth]{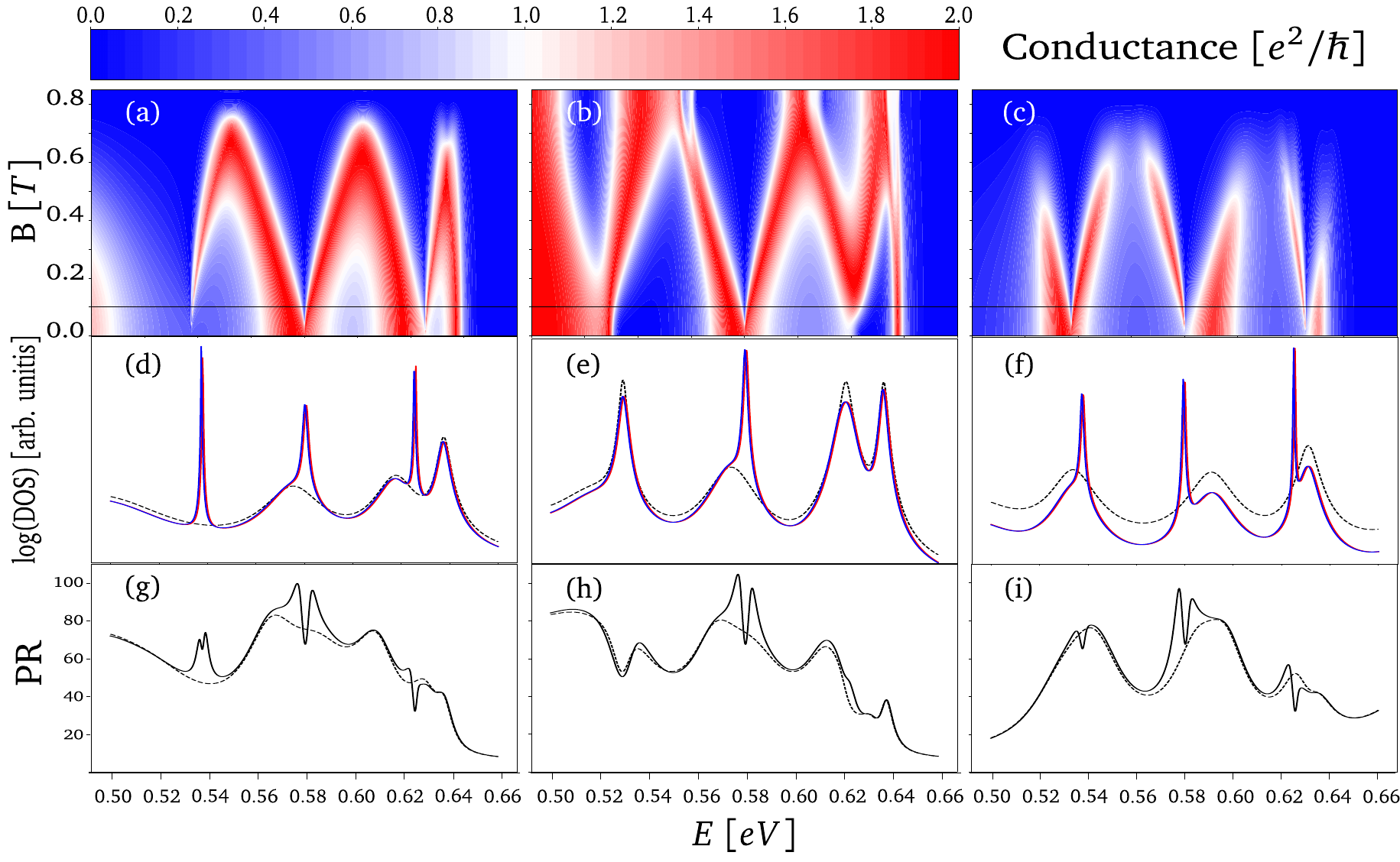}
    \caption{Conductance of the nanowire network against the magnetic field $B$ and the Fermi energy $E$. Panels~(a), (b) and (c) correspond to the structures presented in figure~\ref{fig: 1}(a), (b) and (c), respectively. The horizontal black line indicated the magnetic field at which the density of states and participation ratio are calculated ($B = 0.1\,$T). Panels~(d), (e) and (f) display the density of states with (solid line) and without (dashed line) magnetic field. Red (blue) line corresponds to the density of states with spin up (down). Panels~(g), (h) and (i) represent the participation ratio with (solid line) and without (dashed line) magnetic field for the up spin projection.}
    \label{fig: 2}
\end{figure*}

From figure~\ref{fig: 2}(d), (e), and (f), we can see that all the configurations show a clear difference in the density of states when the magnetic field is applied for both spin projections since marked peaks appear at energies $E_\mathrm{BIC}$. This change is due to the interaction that exists between the BICs and the continuum energy spectrum after applying the magnetic field. As anticipated, the presence of the BICs can be noticed in figures~\ref{fig: 2}(a), (b) and (c), where the BIC at $E_\mathrm{BIC} \approx 0.58\,$eV is the more remarkable one as it is shared among all the configurations, as resonances or antiresonances in the conductance plots. For instance, the DOS around each BIC can be expressed as the superposition of two Lorentzian line shapes centered at energies $\varepsilon_{\pm}$
\begin{equation} 
    \rho (E)\approx \frac{1}{\pi} \left(\frac{\Gamma_{+}}{(E-\varepsilon_{+})^2+\Gamma_{+}^2}+\frac{\Gamma_{-}}{(E-\varepsilon_{-})^2+\Gamma_{-}^2}\right)\ ,
\end{equation}
where $\Gamma_{+}$ and $\Gamma_{-}$ stand for the width of the states strongly and weakly coupled to the continuum, respectively. As $\Gamma_{-} \rightarrow 0$ the DOS can written as
\begin{equation} 
    \rho(E)\approx \frac{1}{\pi} \frac{\Gamma_{+}}{(E-\varepsilon_{+})^2+\Gamma_{+}^2}+\delta (E-\varepsilon_{-})\ .
\end{equation}

The Lorentzian line shapes centered at energy $\varepsilon_{-}$ becomes a Dirac-$\delta$ function, a signature of the BIC. On the other hand, the conductance can be expressed as a convolution of Breit-Wigner and Fano line shapes in the following form
\begin{align}
    G(E)&\approx \frac{\Gamma_{+}^2}{(E-\varepsilon_{+}-\Gamma_{+})^2}\,\frac{|(E-\varepsilon_{-})/\Gamma_{-}+q|^2}{(E-\varepsilon_{-})^2/\Gamma_{-}^2+1}
    \nonumber \\
    & +\frac{\eta^2}{(E-\varepsilon_{-})^2+(\mathrm{Im}(q) \Gamma_{-})^2}\ ,
\end{align}
where the Fano parameter $q$ is a complex number and the parameter $\eta$ is proportional to $\Gamma_{-}$.

\begin{figure*}
    \centering
    \includegraphics[width = 0.9\linewidth]{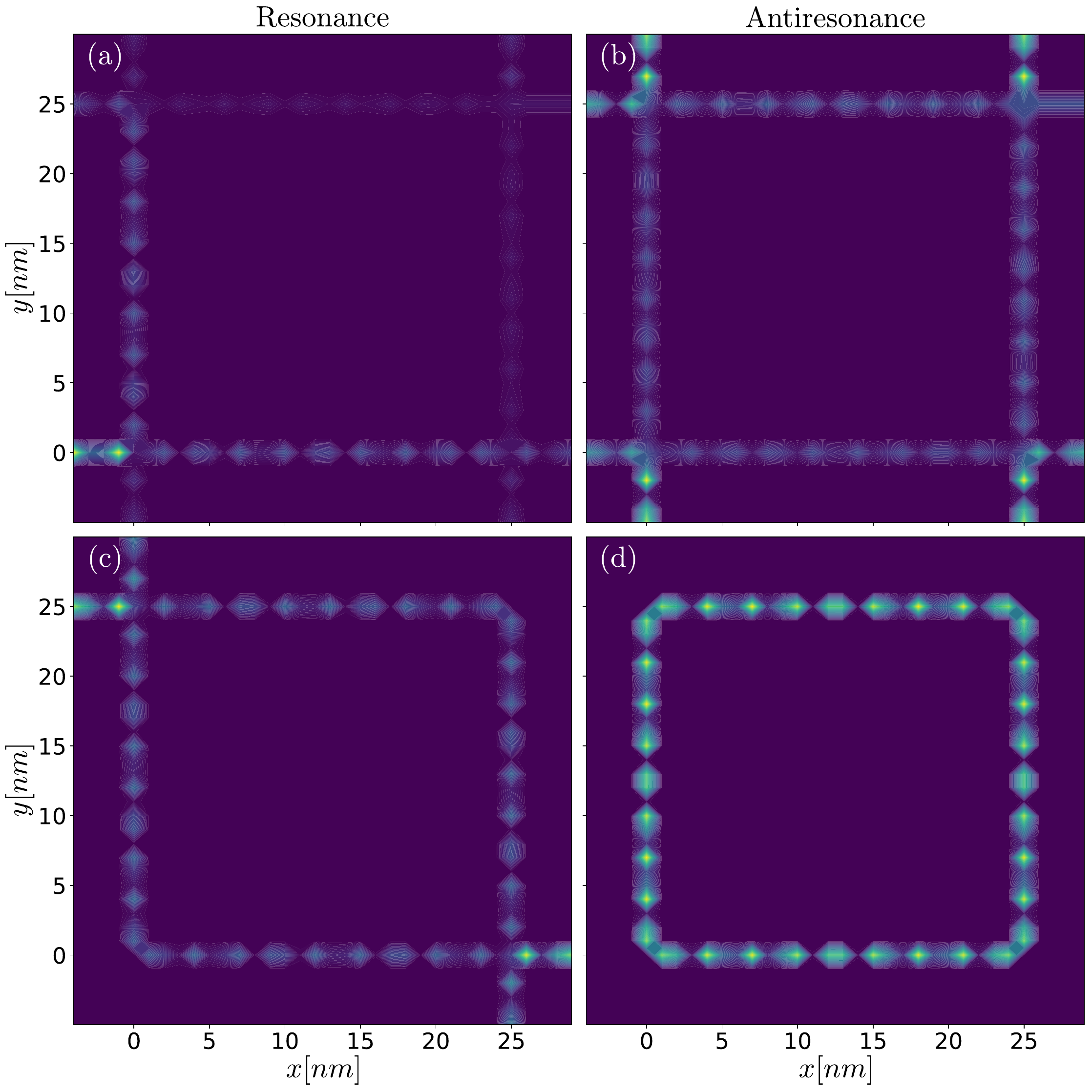}
    \caption{Probability density for the structure shown in figure~\ref{fig: 1}(a). The first (second) row displays the probability density in the absence (presence) of an external magnetic field. The first column shows the probability density at the energy where the resonance in the conductance occurs, $E\approx 0.54\,$eV, while the second column is at $E\approx 0.58\,$eV when there is an anti-resonance in the conductance.}
    \label{fig: 3}
\end{figure*}

To confirm that these peaks correspond to BICs, we have plotted the PR against the Fermi energy with and without magnetic field [figure~\ref{fig: 2}(g), (h), and (i)]. Here, we can see that at the energy of the BICs, $E_\mathrm{BIC}$, the PR is increased (decreased) when a resonance (anti-resonance) occurs in the conductance. This behaviour is related with a delocalization (localization) of the state, as we can see in figure~\ref{fig: 3}. In addition, this behavior is followed by a higher increase of the PR in the vicinity of $E_\mathrm{BIC}$, provoked by mixing the extended states with the BICs.

Although there are BICs in all the presented structure configurations, the first one will be the chosen lead's connection for the next section. This is because it have resonant-like BICs, that will be better resolved by experiments.

\subsection{Impact of the system geometry}

\begin{figure*}
    \centering
    \includegraphics[width = 0.9\linewidth]{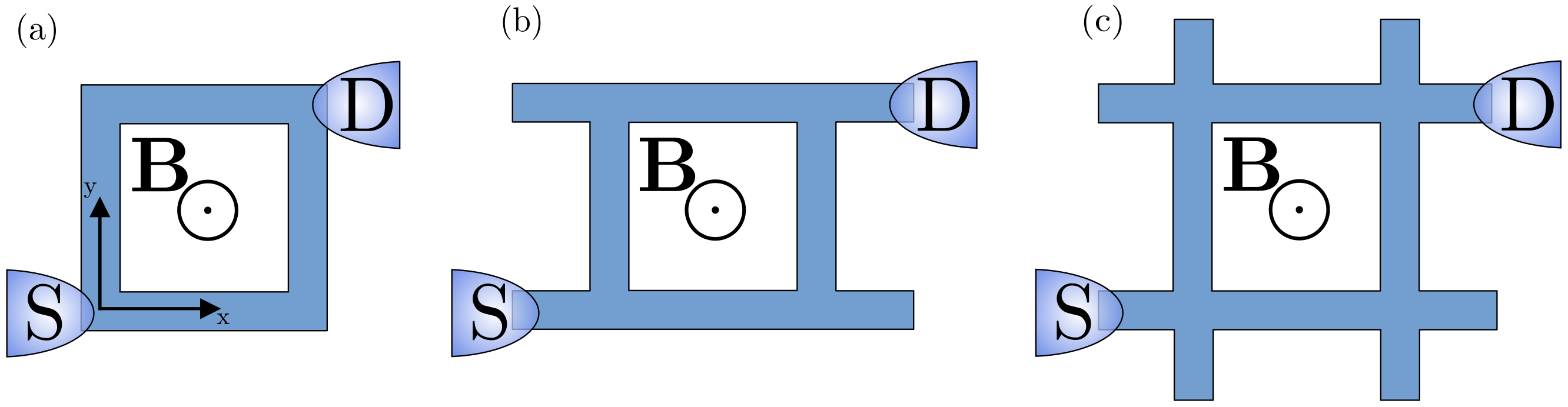}
    \caption{Sketch of the configurationally different geometries of the nanowire network. S and D represents the source and drain leads. All the structures are threaded by a perpendicular magnetic field.}
    \label{fig: 4}
\end{figure*}

\begin{figure*}
    \centering
    \includegraphics[width = 0.9\linewidth]{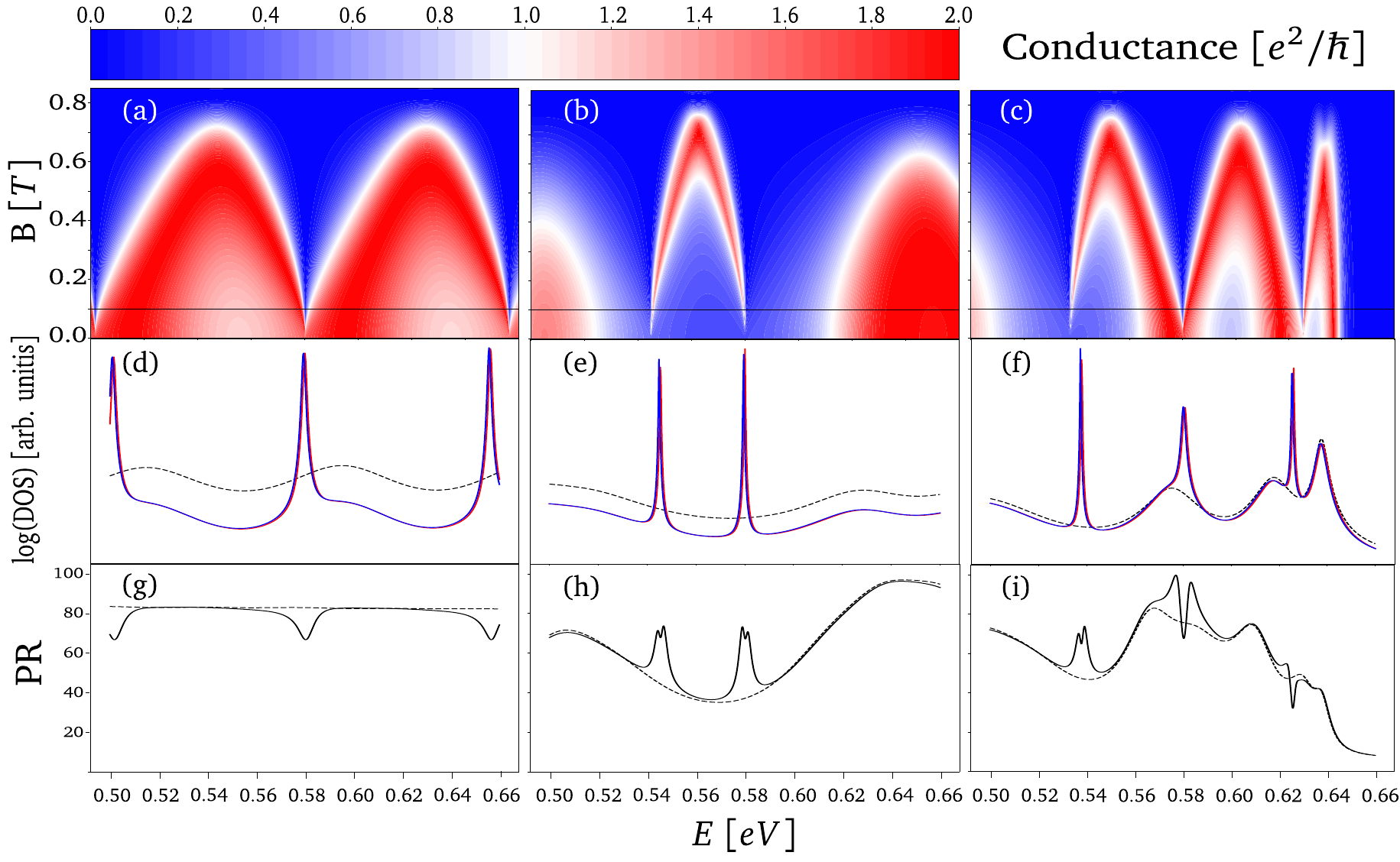}
    \caption{Conductance of the nanowire network against the magnetic field $B$ and the Fermi energy $E$. Panels~(a), (b) and (c) correspond to the structures presented in figure~\ref{fig: 4}(a), (b) and (c), respectively. The horizontal black line indicated the magnetic field at which the density of states and participation ratio are calculated ($B = 0.1\,$T). Panels~(d), (e) and (f) display the density of states with (solid line) and without (dashed line) magnetic field. Red (blue) line corresponds to the density of states with spin up (down). Panels~(g), (h) and (i) represent the participation ratio with (solid line) and without (dashed line) magnetic field for the up spin projection. Panels (c), (f) and (i) present the same data as in figure~\ref{fig: 2}(a), (e) and (h) to facilitate the comparison between the structures.}
    \label{fig: 5}
\end{figure*}

We will consider three different system geometries. The first geometry will have no dead-end chains resembling a square. In the second configuration, the scattering region will have only horizontal dead-end chains in each square corner. Finally, the last one will have either horizontal and vertical dead-end chains in each corner of the square (see figure~\ref{fig: 4}). From the density of states shown in figure~\ref{fig: 5}, we realize that BICs emerge in all the considered geometries, either as resonances or anti-resonances. BICs are better revealed in an energy window where the conductance is zero in the absence of an applied magnetic field, and some resonances appear after its application; these resonances will correspond to the BICs. Therefore, the most favorable configuration to detect the occurrence of BICs is the structure shown in figure~\ref{fig: 4}(b). Moreover, it is noticeable that the BIC at $E_\mathrm{BIC} \approx 0.58\,$eV is again shared in all the structures. For this reason, this will be the geometry and energy studied in the following section. 




\subsection{Impact of the Rashba SOI}

\begin{figure*}
    \centering
    \includegraphics[width = 0.9\linewidth]{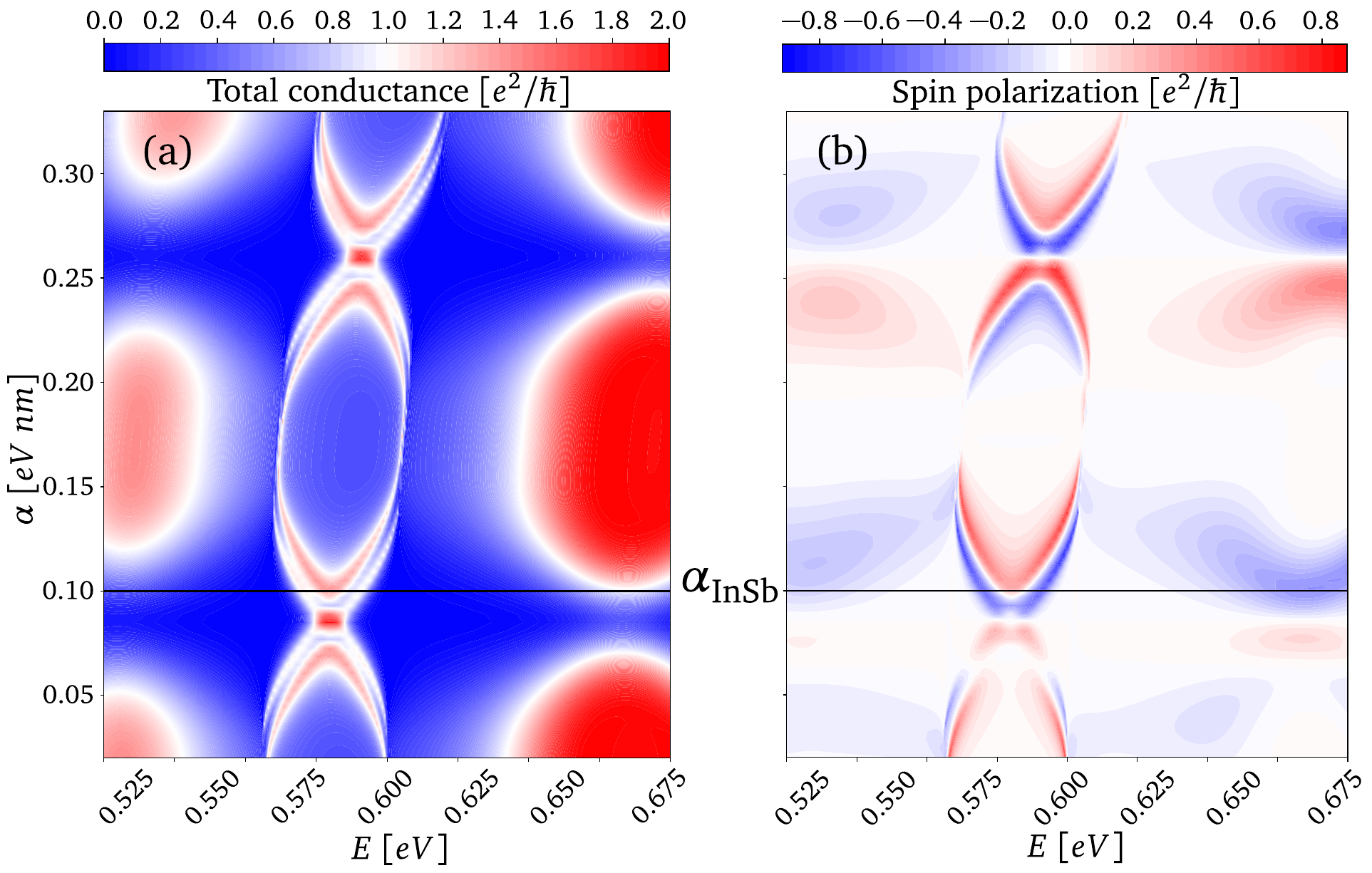}
    \caption{(a)~Conductance and (b)~spin-dependent conductance of the nanowire network shown in figure~\ref{fig: 4} (b) against the Rashba SOI $\alpha$ and the Fermi energy $E$ when a perpendicular magnetic field is applied ($B = 0.1\,$T).  Here, positive values on the spin conductances mean that the up spin projection dominates the conductance. The black horizontal line represents the Rashba SOI estimated for InSb nanowires.}
    \label{fig: 6}
\end{figure*}

In previous calculations, we have neglected the Rashba SOI  ($\alpha = 0$). In figure~\ref{fig: 6}, we show the evolution of the BICs against the magnitude of the Rashba SOI and the Fermi energy when a magnetic field is applied. This figure shows that the Rashba SOI smears out the resonances, making the BICs harder to identify. In addition, we can see an oscillatory behavior on increasing $\alpha$, associated with the Aharonov-Casher effect~\cite{Konig2006}. InSb nanowires have a Rashba SOI of the order of $\alpha_{InSb} = 0.1\,$eV~\cite{Weperen2015}, which is not the optimum value for observing the BICs. For this reason, an external electric field might be necessary to change $\alpha$, therefore facilitating the observation of the BICs. Additionally, we can see that the nanowire network can be used in spintronics because it displays a non-zero spin conductance at some specific energies. Also, with the applied electric field, we can control which spin projection will be dominant, allowing us to use them as spin filters.

\section{Conclusions}

In this article, we have proposed a setup that could be used to search for BICs in electronic experiments. We have shown that the hashtag nanowire networks will reveal the BICs as resonances or antiresonances of the conductance by varying the magnetic field and the Fermi energy. This statement is supported by calculating the density of states whose peaks associated with BICs broaden upon rising the magnetic field, i.e., increasing the coupling to the extended states of the continuum energy spectrum. In addition, we have studied different different configurations to elucidate the best one for uncovering BICs. The study showed that the optimum geometry is the one with horizontal dead-end chains connected to each corner and the leads connected horizontally to opposite corners. Finally, we studied the impact of the Rashba SOI on the BICs. We found that this interaction blurs the BICs, making them harder to find. However, it opens a way to use hashtag-like nanowire networks in spintronics as spin filters while tuning the Rashba SOI by strain or applied electric field.
 
\acknowledgments

We acknowledge financial support from Comunidad de Madrid (Recovery, Transformation and Resilience Plan) and NextGenerationEU from the European Union (Grant MAD2D-CM-UCM5), Agencia Estatal de Investigación of Spain (Grant PID2022-136285NB-C31/2), Junta de Castilla y León  and The European Regional Development Fund (SA106P23) and FONDECYT (Grants 1220700 and 1201876).

\bibliography{references}


\end{document}


\begin{center}
    \vspace{16mm}
	\textsc{\Large \textbf{Suplementary material\\ Uncovering bound states in the continuum in InSb nanowire networks}}
\end{center}

\section{Surface Green's function}

In this section, an analytical expression of the surface Green's function for the leads and chains will be obtained by means of the Dyson equation
\begin{equation}
    \hat{G} = \hat{g} + \hat{g}\hat{v}\hat{G}
\end{equation}
where $\hat{g}$ are known Green's functions and $\hat{v}$ is the coupling matrix. For both the chains and leads, $\hat{g} = g_0$ will be taken as the Green's function of a single quantum dot (QD).

\subsection{Surface Green's function of the lead}

To find the surface Green's function of one of the  leads, we will consider it as a semi-infinite chain where we add a new site
\begin{equation*}
    G_{00} = -g_0 v_l G_{10}, \hspace{0.2cm} G_{10} = -g_l v_l G_{00}\ .
\end{equation*}

Here, $v_l$ and $g_l$ are the hopping energy and surface Green's function of the lead respectively. Since the lead is semi-infinite, its surface Green's function should not change with the addition of a single site. Therefore $G_{00} = g_l$ and thus we find
\begin{equation}
    g_l = \frac{1- \sqrt{1 - v_l^2 g_0^2}}{2v^2 g_0} = \frac{e^{-ik_l}}{v_l}\ .
\end{equation}

\subsection{Surface Green's function of a finite chain}

We will consider the chain as a finite array of $N$ QDs coupled between them. Additionally, we will consider that the hopping parameter includes a magnetic phase due to the presence of a magnetic field. Using the Dyson equation we find
\begin{equation*}
    g_{11} = g_0 - g_0 v e^{-\mathrm{i} \phi} g_{21}, \hspace{0.2cm} g_{21} = -g_0 v (e^{\mathrm{i} \phi} g_{11} + e^{-\mathrm{i} \phi} g_{31}), \hspace{0.2cm} \cdots \hspace{0.2cm} , \hspace{0.2cm} g_{N1} = -g_0 v e^{\mathrm{i} \phi} g_{N-1\, 1}\ ,
\end{equation*}
where $\phi = e B y \Delta x/\hbar$ is the magnetic phase resulting from the Peierls substitution considering the gauge  $\boldsymbol{A} = (-By, 0, 0)$ and $v$ is the hopping parameter of the chain. Solving this system of equations we get
\begin{equation}
    g_{11} = \frac{1}{v} \frac{D_{N-1}}{D_N}, \hspace{0.2cm} g_{N1} = \frac{(-1)^{N-1}}{v} \frac{e^{\mathrm{i}(N-1)\phi}}{D_N}\ ,
\end{equation}
where $D_N = \sin [(N+1)\theta]/\sin \theta$ and $\theta = (v g_0)^{-1}$. These calculations have considered the chain in the $x$ direction.

\section{Transmission coefficient in terms of the surface Green's function}
\begin{figure}
    \centering
    \includegraphics[width = 0.8\linewidth]{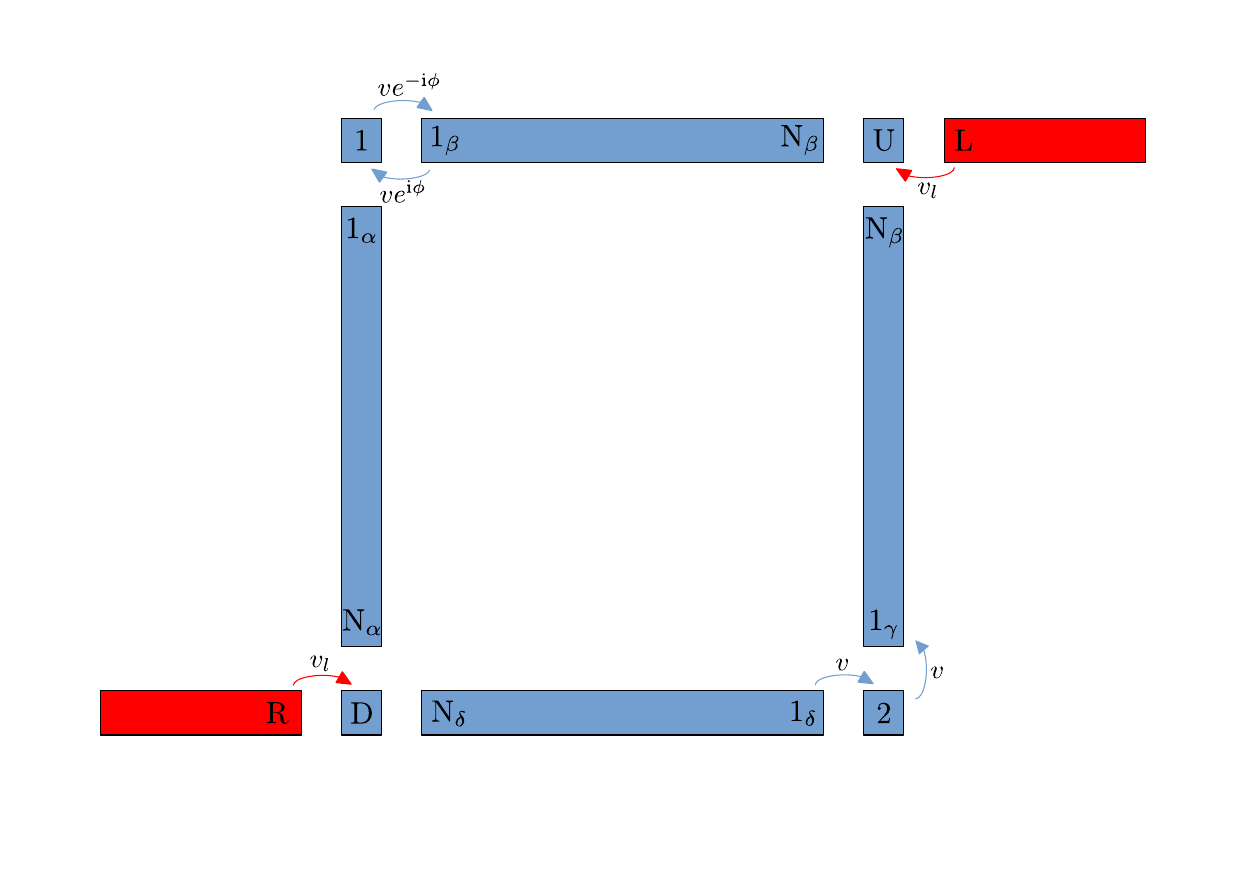}
    \caption{Sketch of the system. The red rectangles correspond to the left and right leads while the blue region is the scattering region. The coupling between leads and the center region is considered to be the same as in the lead and its represented by red arrows while the hopping in the center region is represented by blue arrows.}
    \label{fig: 1}
\end{figure}
In this section, the transmission coefficient is calculated for the simplest structure considered in the main text, namely the hashtag configuration without dead-end chains. The method we used is named the surface Green's functions matching formalism (see \href{https://journals.aps.org/prb/abstract/10.1103/PhysRevB.60.7828}{10.1103/PhysRevB.60.7828}). We split the system into three parts: two semi-infinite chains acting as ideal leads and the scattering region, which is described as four coupled QDs located at the corners (see figure \ref{fig: 1}). For simplicity, we will neglect Zeeman splitting in the following calculations since the magnetic fields considered are weak.

First, we use the Lippmann-Schwinger equation for the coupling between the leads and the scattering region
\begin{equation}
    \left| \psi \right> = \left| \varphi \right> + \hat{G} \hat{V} \left| \varphi \right>\ ,
\end{equation}
where $\left| \psi \right>$ and $\hat{G}$ are the wave function and Green's function of the coupled system respectively, $\left| \varphi \right>$ is the wave function of the source lead ($L$) and $\hat{V}$ is the coupling matrix between the leads and the scattering region (red arrows), which is zero everywhere except at the top-right and bottom-left corners. The transmission coefficient reads
\begin{equation*}
    t = 2\mathrm{i} \sin(k_l)e^{-\mathrm{i} k_l} v_l \Tilde{G}_{RD}\ ,
\end{equation*}
where $k_l$ is the crystal momentum in the leads.

Now we need to find the dressed Green's function in terms of those of the scattering region. To this end, we will use the Dyson equation considering the upper and lower parts of the structure connecting the QDs as known magnitudes. Hence
\begin{equation*}
\begin{array}{cc}
     \Tilde{G}_{RD} = -g_l v_l \Tilde{G}_{UD}\ , 
     \hspace{0.5cm}  
     \Tilde{G}_{UD} = -g_0 v (e^{\mathrm{i}\phi}\Tilde{G}_{N_\beta D} +  \Tilde{G}_{N_\gamma D}) -g_0 v_l \Tilde{G}_{R D}\ , 
     \\\\
     \Tilde{G}_{N_\beta D} = -\Tilde{G}^0_{N_\beta N_\beta} v e^{-\mathrm{i} \phi} \Tilde{G}_{U D} - \Tilde{G}_{N_\beta N_\alpha}^0 v \Tilde{G}_{D D}\ , 
     \hspace{0.5cm}  
     \Tilde{G}_{N_\gamma D} = -\Tilde{G}^0_{N_\gamma N_\gamma} v \Tilde{G}_{U D} - \Tilde{G}_{N_\gamma N_\delta}^0 v \Tilde{G}_{D D}\ , 
\end{array}
\end{equation*}
\begin{equation*}
\begin{array}{cc}
     \Tilde{G}_{DD} = g_0 -g_0 v (\Tilde{G}_{N_\delta D} +  \Tilde{G}_{N_\alpha D}) - g_0 v_l \Tilde{G}_{L D}\ , 
     \\\\
     \Tilde{G}_{N_\delta D} = -\Tilde{G}_{N_\delta N_\gamma}^0 v \Tilde{G}_{U D} - \Tilde{G}^0_{N_\delta N_\delta} v \Tilde{G}_{D D}\ , \hspace{0.5cm}  
     \Tilde{G}_{N_\alpha D} = -\Tilde{G}_{N_\alpha N_\beta}^0 v \Tilde{G}_{U D} - \Tilde{G}^0_{N_\alpha N_\alpha} v \Tilde{G}_{D D}\ ,
\end{array}
\end{equation*}

After that, we need to find the Green's function of the different branches ($\Tilde{G}^0$) connecting the QD using the same method.
\begin{equation*}
    \begin{array}{cc}
         \Tilde{G}^0_{N_\alpha N_\alpha} = g_{N_\alpha N_\alpha} - g_{N_\alpha 1_\alpha}v \Tilde{G}^0_{1 N_\alpha}\ , 
         \hspace{0.5cm}  
         \Tilde{G}^0_{N_\beta N_\beta} = g_{N_\beta N_\beta} - g_{N_\beta 1_\beta}v e^{\mathrm{i}\phi} \Tilde{G}^0_{1 N_\beta}\ ,
         \\\\ 
         \Tilde{G}^0_{N_\gamma N_\gamma} = g_{N_\gamma N_\gamma} - g_{N_\gamma 1_\gamma}v \Tilde{G}^0_{2 N_\gamma}\ , 
         \hspace{0.5cm}
         \Tilde{G}^0_{N_\delta N_\delta} = g_{N_\delta N_\delta} - g_{N_\delta 1_\delta}v \Tilde{G}^0_{2 N_\delta}\ , 
         \\\\
         \Tilde{G}^0_{1 N_\alpha} = - g_0 v (e^{-\mathrm{i}\phi} \Tilde{G}^0_{1_\beta N_\alpha} + \Tilde{G}^0_{1_\alpha N_\alpha})\ ,\hspace{0.5cm} 
         \Tilde{G}^0_{1 N_\beta} = - g_0 v (e^{-\mathrm{i}\phi} \Tilde{G}^0_{1_\beta N_\beta} + \Tilde{G}^0_{1_\alpha N_\beta})
         \\\\
         \Tilde{G}^0_{2 N_\gamma} = - g_0 v (\Tilde{G}^0_{1_\gamma N_\gamma} + \Tilde{G}^0_{1_\delta N_\gamma})\ ,
         \hspace{0.5cm}
         \Tilde{G}^0_{2 N_\delta} = - g_0 v (\Tilde{G}^0_{1_\gamma N_\delta} + \Tilde{G}^0_{1_\delta N_\delta})        
        \\\\
        \Tilde{G}^0_{1_\alpha N_\alpha} = g_{1_\alpha N_\alpha} - g_{1_\alpha 1_\alpha}v \Tilde{G}^0_{1 N_\alpha}\ ,
         \hspace{0.5cm}
         \Tilde{G}^0_{1_\beta N_\beta} = g_{1_\beta N_\beta} - g_{1_\beta 1_\beta}v e^{\mathrm{i}\phi} \Tilde{G}^0_{1 N_\beta}\ ,
        \\\\
         \Tilde{G}^0_{1_\gamma N_\gamma} = g_{1_\gamma N_\gamma} - g_{1_\gamma 1_\gamma}v \Tilde{G}^0_{2 N_\gamma}\ , 
         \hspace{0.5cm}
         \Tilde{G}^0_{1_\delta N_\delta} = g_{1_\delta N_\delta} - g_{1_\delta 1_\delta}v \Tilde{G}^0_{2 N_\delta}\ , 
         \\\\
         \Tilde{G}^0_{1_\alpha N_\alpha} = - g_{1_\alpha 1_\alpha}v \Tilde{G}^0_{1 N_\beta}\ ,
         \hspace{0.5cm}
         \Tilde{G}^0_{1_\beta N_\beta} = - g_{1_\beta 1_\beta}v e^{\mathrm{i}\phi} \Tilde{G}^0_{1 N_\alpha}\ ,
         \\\\
         \Tilde{G}^0_{1_\gamma N_\delta} = - g_{1_\gamma 1_\gamma}v \Tilde{G}^0_{2 N_\delta}\ , 
         \hspace{0.5cm}
         \Tilde{G}^0_{1_\delta N_\gamma} = - g_{1_\delta 1_\delta}v \Tilde{G}^0_{2 N_\gamma}\ , 
         \\\\
         \Tilde{G}^0_{N_\alpha N_\beta} = - g_{N_\alpha 1_\alpha}v \Tilde{G}^0_{1 N_\beta}\ , 
         \hspace{0.5cm}  
         \Tilde{G}^0_{N_\beta N_\alpha} = - g_{N_\beta 1_\beta}v e^{\mathrm{i}\phi} \Tilde{G}^0_{1 N_\alpha}\ ,
         \\\\ 
         \Tilde{G}^0_{N_\gamma N_\delta} = - g_{N_\gamma 1_\gamma}v \Tilde{G}^0_{2 N_\delta}\ , 
         \hspace{0.5cm}
         \Tilde{G}^0_{N_\delta N_\gamma} = - g_{N_\delta 1_\delta}v \Tilde{G}^0_{2 N_\gamma}\ .
    \end{array}
\end{equation*}
Closing these equations and inserting the solution into the previous expressions, also considering that all chains have the same number of sites $N$, we can find the transmission amplitude
\begin{equation}
    t = \frac{4\mathrm{i} \sin(k_l) e^{-\mathrm{i}k_l} g_l v_l^2 v^2 g_0^2 \Tilde{G}_N e^{\mathrm{i} (N+1) \phi / 2} \cos [(N+1) \phi / 2]}{1 - 2g_0(vt^2\Tilde{G} + t_l^2g_l) +  g_0^2[g_l^4v_l^2 + 4v^2 v_l^2 g_l \Tilde{G} + 2v^4(2\Tilde{G} - \Tilde{G}_N^2 - \Tilde{G}_N^2\cos[(N+1)\phi]]}  \ , 
    \label{eq: t}
\end{equation}
where $\Tilde{G} = \Tilde{G}^0_{N_\alpha N_\alpha}$, $\Tilde{G}_N = \Tilde{G}^0_{N_\gamma N_\delta}$, $\Tilde{G}_{N_i N_j}^0 = (\Tilde{G}_{N_j N_i}^0)^*$ and $\Tilde{G}_{N_\alpha N_\beta}^0 = \Tilde{G}_{N_\gamma N_\delta}^0 e^{-\mathrm{i}  N \phi}$. We can use this expression to determine how the conductance will behave when a magnetic field is applied using the Landauer-B\"{u}ttiker formalism. Thus, the chosen parameters for the calculation are the same as the ones in the paper i.e. $m^*_\mathrm{InSb} = 0.014\,m_e$, $a = 2\,$nm and $L = 50\,$nm.
\begin{figure}
    \centering
    \includegraphics[width = \linewidth]{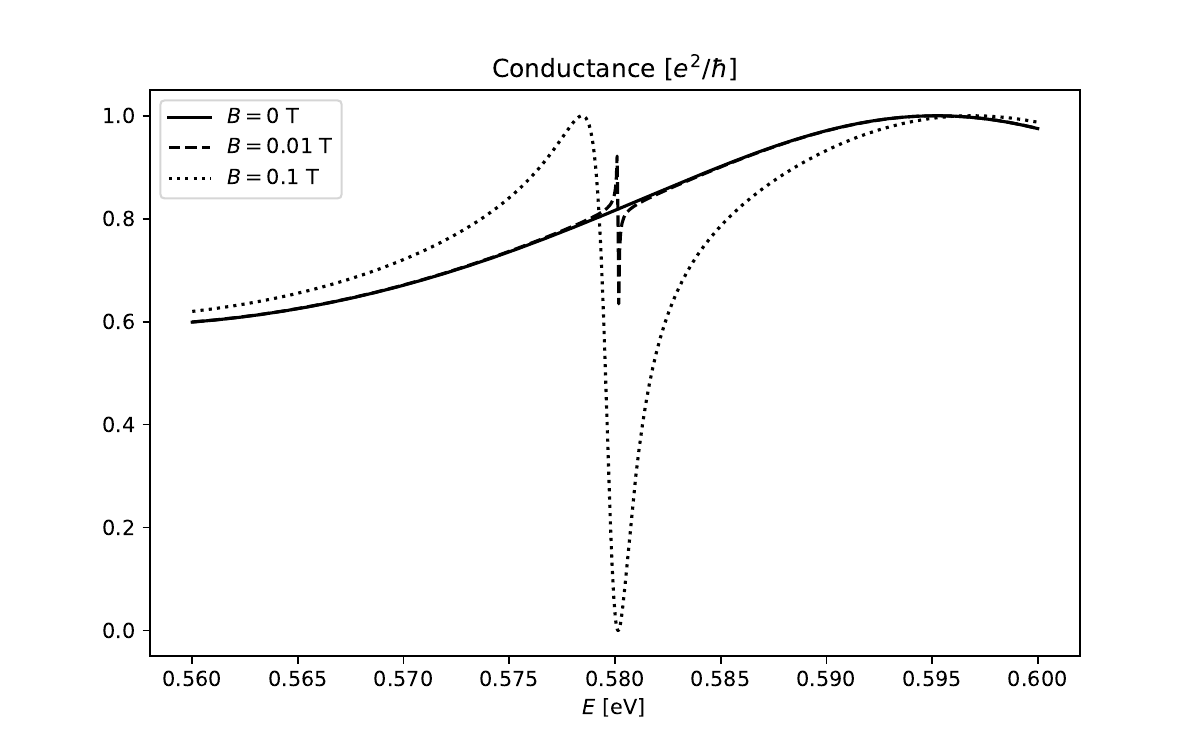}
    \caption{Conductance of the system. The different line styles correspond to different magnetic applied}
    \label{fig: 2}
\end{figure}

As we state in the paper, we can see from figure \ref{fig: 2} that as long as the magnetic field is higher, the resonance becomes broader. In this example, the BIC induces an antiresonance in the conductance, but the same trend is expected if the BIC induces a resonance.